\documentclass[12pt]{article}
\usepackage{euscript}
\usepackage{amssymb}
\usepackage{amsmath}
\usepackage{graphicx}
\usepackage[small,sc,center]{titlesec}
\usepackage{indentfirst}

\newcommand{\nc}{\newcommand}
\nc{\pd}[2]{\dfrac{\partial{#1}}{\partial{#2}}}
\nc{\pdd}[3]{\left(\pd{#1}{#2}\right)_{#3}}
\nc{\alphac}{\alpha_\mathrm{c}}
\nc{\dt}    {D_\mathrm{t}}
\nc{\et}    {E_\mathrm{t}}
\nc{\flux}  {\mathbf{F}}
\nc{\K}     {\mathrm{K}}
\nc{\mzams} {M_\mathrm{ZAMS}}
\nc{\pt}    {P_\mathrm{t}}
\nc{\pv}    {P_\mathrm{v}}
\nc{\st}    {S_\mathrm{t}}
\nc{\Teff}  {T_\mathrm{eff}}
\nc{\vc}    {v_\mathrm{c}}
\nc{\Xs}    {X_\mathrm{s}}

\begin{document}

\begin{center}
\textbf{PULSATIONAL INSTABILITY OF YELLOW HYPERGIANTS}

\textbf{Yu. A. Fadeyev\footnote{E--mail: fadeyev@inasan.ru}}

\textit{Institute of Astronomy, Russian Academy of Sciences, Pyatnitskaya ul. 48, Moscow, 109017 Russia}

Received

\end{center}

\textbf{Abstract} --- Instability of population~I ($X=0.7$, $Z=0.02$) massive stars
against radial oscillations during the post--main--sequence gravitational contraction
of the helium core is investigated.
Initial stellar masses are in the range $65M_\odot\le\mzams\le 90M_\odot$.
In hydrodynamic computations of self--exciting stellar oscillations we assumed
that energy transfer in the envelope of the pulsating star is due to radiative heat
conduction and convection.
The convective heat transfer was treated in the framework of the theory of
time--dependent turbulent convection.
During evolutionary expansion of outer layers after hydrogen exhaustion in the stellar core
the star is shown to be unstable against radial oscillations while its
effective temperature is $\Teff > 6700$K for $\mzams=65M_\odot$ and
$\Teff > 7200$K for $\mzams=90M_\odot$.
Pulsational instability is due to the $\kappa$--mechanism in helium ionization zones
and at lower effective temperature oscillations decay because of
significantly increasing convection.
The upper limit of the period of radial pulsations on this stage of evolution
does not exceed $\approx 200$~day.
Radial oscillations of the hypergiant resume during evolutionary contraction
of outer layers when the effective temperature is
$\Teff > 7300$K for $\mzams=65M_\odot$
and $\Teff > 7600$K for $\mzams=90M_\odot$.
Initially radial oscillations are due to instability of the first overtone
and transition to fundamental mode pulsations takes place at higher
effective temperatures
($\Teff > 7700$K for $\mzams=65M_\odot$ and $\Teff > 8200$K for $\mzams=90M_\odot$).
The upper limit of the period of radial oscillations
of evolving blueward yellow hypergiants
does not exceed $\approx 130$~day.
Thus, yellow hypergiants are stable against radial stellar pulsations
during the major part of their evolutionary stage.

Keywords: \textit{stars: variable and peculiar}.

\newpage
\section{INTRODUCTION}

More than a half--century ago Feast and Thackeray (1956) pointed out
the significant scatter in luminosities (up to 7 magnitudes) of supergiants
of the Large Magellanic Cloud.
The most luminous stars of this sample of spectral types F and G
the authors of this work named as super--supergiants.
Later the name super--supergiant was changed to a hypergiant and
the luminosity class of such objects was denoted as Ia${}^+$ or Ia0
(de Jager 1980).
The high luminosity and the low surface gravity favour the strong stellar wind,
so that as one more criterion of the hypergiant one should consider the widely
broadened emission and absorption lines.
Evolutionary status of yellow hypergiants are thought to be that they are
the post--main--sequence massive stars (i.e. with initial masses
$\mzams > 40M_\odot$) undergoing gravitational contraction of the helium core
(Meynet et al. 1994; Stothers and Chin 1996; Langer et al. 2007).

Yellow hypergiants are variable objects but the common point of view
on nature of their variability does not exist yet.
Stothers and Chin (1996, 2001) showed that outer layers of hypergiants might
be dynamically unstable because the adiabatic exponent becomes smaller
its critical value ($\Gamma_1 < 4/3$) due to partial ionization of helium
and the large contribution of radiation pressure.
Outer layers of the hypergiant can be unstable also due to convection in
the hydrogen and helium ionization zones.
De Jager and Nieuwenhuijzen(1997) and de Jager et al. (2001)
estimated that the gradient of the turbulent pressure in subphotospheric layers
might reduce the effective gravity almost to zero.
However this conclusion was not confirmed by Stothers (2003) who showed that
the turbulent pressure is insignificant and this effect can be neglected.

Together with irregular enhancements of the stellar wind some yellow hypergiants show
cyclic light variations.
The most remarkable example is $\rho$~Cas with light amplitude of a few tenths of
a magnitude and the period of order of several hundred
days (Arellano Ferro 1985; Sheffer and Lambert 1986; Zsoldos and Percy 1991;
Lobel et al. 1994, 1998).
Most of these authors mentioned the fact that cyclic light variations
of $\rho$~Cas are very similar to those of pulsating stars.
Moreover, near--IR CO line profiles demonstrate splitting similar to that
observed in pulsating supergiants of intermediate spectral types such as
RV~Tau and R~CrB variables (Gorlova et al. 2006).
At the same time there seems to be a contradiction with pulsation hypothesis
because light variations with period of $\approx 500$~day cannot be explained
in terms of radial stellar pulsations for typical masses and radii of
yellow hypergiants (Sheffer and Lambert 1986).
Here one should be noted that no studies of pulsational instability of
yellow hypergiants have been done yet, so that disussion of their
photometric variability in terms of stellar pulsations remains hypothetical.

As shown in our previous paper (Fadeyev 2011) the population~I stars
($Z\ge 0.02$) with initial mass $\mzams > 60M_\odot$ become unstable against
radial oscillations during the main--sequence evolutionary stage and
they pulsate at least up to hydrogen exhaustion in the stellar core.
The goal of the present study is to consider the pulsation properties
of massive stars during the later evolutionary stage of gravitational contraction
of the helium core and to determine the role of radial stellar
pulsations in variability of yellow hypergiants.

\section{THE METHOD OF COMPUTATIONS}

The methods used to compute the stellar evolution and self--exciting radial stellar
oscillations were described in our previous papers
(Fadeyev 2007, 2008a, 2008b, 2010, 2011),
so that below we outline the most important improvements employed in hydrodynamic
calculations of the present work.
In studies of radial pulsations of Wolf--Rayet stars (Fadeyev 2007, 2008a, 2008b)
and LBV--stars (Fadeyev 2010) the hydrodynamic computations were done in assumption
of radiation transfer without convection.
Bearing in mind complete ionization of hydrogen and helium in outer layers of these stars
such an approach seems to be appropriate.
In calculations of nonlinear pulsations of massive main--sequense stars with
effective temperatures $\Teff > 2\times 10^4$K (Fadeyev 2011)
the energy transport was assumed to be due to radiation and convection.
The region of convective heat transfer encompasses the zone of the
iron Z--bump ($T\sim 2\times 10^5$K) and convection was treated in the framework
of the local steady--state mixing length theory
(B\"ohm--Vitense 1958).
However in outer layers (i.e. in hydrogen and helium ionization zones) of yellow hypergiants
with effective temperatures $\Teff < 10^4$K the significant fraction of energy
is transported by convection and interaction of convective elements
with pulsation motions should be taken into account
because of the large pulsation amplitude.
To this end in the present study we employ the equations of time--dependent turbulent
convection obtained by Kuhfuss (1986) from Navier--Stokes equations written for spherical
geometry.

For spherically--symmetric motions the Lagrangean mass coordinate $M_r$ and the
radius $r$ are related by continuity equation $dM_r = 4\pi r^2\rho dr$, where
$\rho$ is the gas density.
Equation of motion describing the radial gas flow velocity $U$ as a function of
time $t$ is
\begin{equation}
\label{motcon}
\pd{U}{t} = - 4\pi r^2 \pd{\left(P + \pt + \pv\right)}{M_r} - \frac{G M_r}{r^2} ,
\end{equation}
where $P$ is the total thermodynamic (gas plus radiation) pressure,
$\pt$ is the turbulent pressure,
$\pv$ is the turbulent viscous pressure,
$G$ is the Newtonian graviational constant.
The energy conservation equation is
\begin{equation}
\label{enereq1}
\pd{\left(E + \et\right)}{t} + \left(P + \pt + \pv\right) \pd{}{t}\left(\frac{1}{\rho}\right) = - \frac{1}{\rho}\nabla\cdot\flux ,
\end{equation}
where $E$ is the specific internal energy (the sum of the translational,
excitation, ionization and radiation energies per gram of material),
$\et$ is the root--mean specific turbulent energy,
$\flux = \flux_\mathrm{rad} + \flux_\mathrm{c} + \flux_\mathrm{t}$ is the total flux,
that is the sum of the radiative, convective (enthalpy) and
turbulent (kinetic energy of convective elements) fluxes.

Additionally we write the equation for conservation of the turbulent energy
\begin{equation}
\label{enereq2}
\pd{\et}{t} + \bigl(\pt + \pv\bigr) \pd{}{t}\left(\frac{1}{\rho}\right) = \st - \dt - \frac{1}{\rho}\nabla\cdot\flux_\mathrm{t} ,
\end{equation}
where $\dt$ is dissipation of turbulent energy due to molecular viscosity
and $\st$ is the source or sink of the turbulent energy due to boyancy forces.
In the isotropic medium the turbulent pressure $\pt$, the mean turbulent energy $\et$
and the root--mean speed of convective elements $\vc$ are related by
\begin{equation}
\pt = \frac{2}{3}\rho\et = \rho\vc^2 .
\end{equation}
Expressions for other quantities in equations (\ref{motcon}) -- (\ref{enereq2})
can be found in papers by Wuchterl and Feuchtinger (1998), Olivier and Wood (2005),
Smolec and Moskalik (2008).

The numerical solution of equations (\ref{motcon}) -- (\ref{enereq2}) was done
with difference equations of the second--order accuracy in both the spatial coordinate $M_r$
and the time $t$.
The equation of motion (\ref{motcon}) was solved with the explicit finite--difference
method,
whereas equations (\ref{enereq1}) and (\ref{enereq2}) were treated implicitly
using the Crank--Nicholson scheme (see, e.g., Richtmyer and Morton 1967).
Thus, at each step of integration with respect to time $t$
the temperature $T$ and the specific turbulent energy $\et$ are determined for all
Lagrangean mass zones of the hydrodynamic model from iterative solution
of linearized difference energy equations.

For the inner boundary condition we employed the assumption of the rigid sphere with
the time--independent flux:
\begin{equation}
\pd{r_0}{t} = \pd{L_0}{t} = 0,
\end{equation}
where $r_0$ and $L_0$ are the radius and luminosity at the inner boundary.
The gas temperature at the inner boundary is $T < 10^6$K and therefore
the thermonuclear energy generation within the pulsating envelope can be
neglected.
In hydrodynamic calculations the luminosity at the inner boundary $L_0$
was treated as the initial parameter of the hydrodynamic model
determined from computations of stellar evolution.
It should be noted that for the sake of stability of the numerical solution
the inner boundary condition of the energy equation was written in assumption
of radiation heat conduction.
Therefore, the inner boundary of the hydrodynamical model is always below the
iron Z--bump convection zone ($T\sim 2\times 10^5$K).
For models considered in the present study the radius of the inner boundary is
$r_0\lesssim 3\times 10^{-2}R$, where $R$ is the equilibrium radius of the
outer boundary.

At the evolutionary stage of the yellow hypergiant the stellar mass is about a half
of its initial mass and in the outer layers of the star there is the negative gradient
of the mean molecular weight.
The spatial inhomogeneity of the chemical composition is due to diminution of the
both stellar mass and mass of convective core during the main--sequence hydrogen burning.
However the mass of outer pulsating layers of the hypergiant is enough small
in comparison with stellar mass ($M_\mathrm{env} < 10^{-3}M$),
so that effects of inhomogeneity of chemical composition with respect to
spatial coordinate remain always negligible.

\section{RESULTS OF COMPUTATIONS}

\subsection{Stellar evolution}

\begin{figure}
\centerline{\includegraphics[width=15cm]{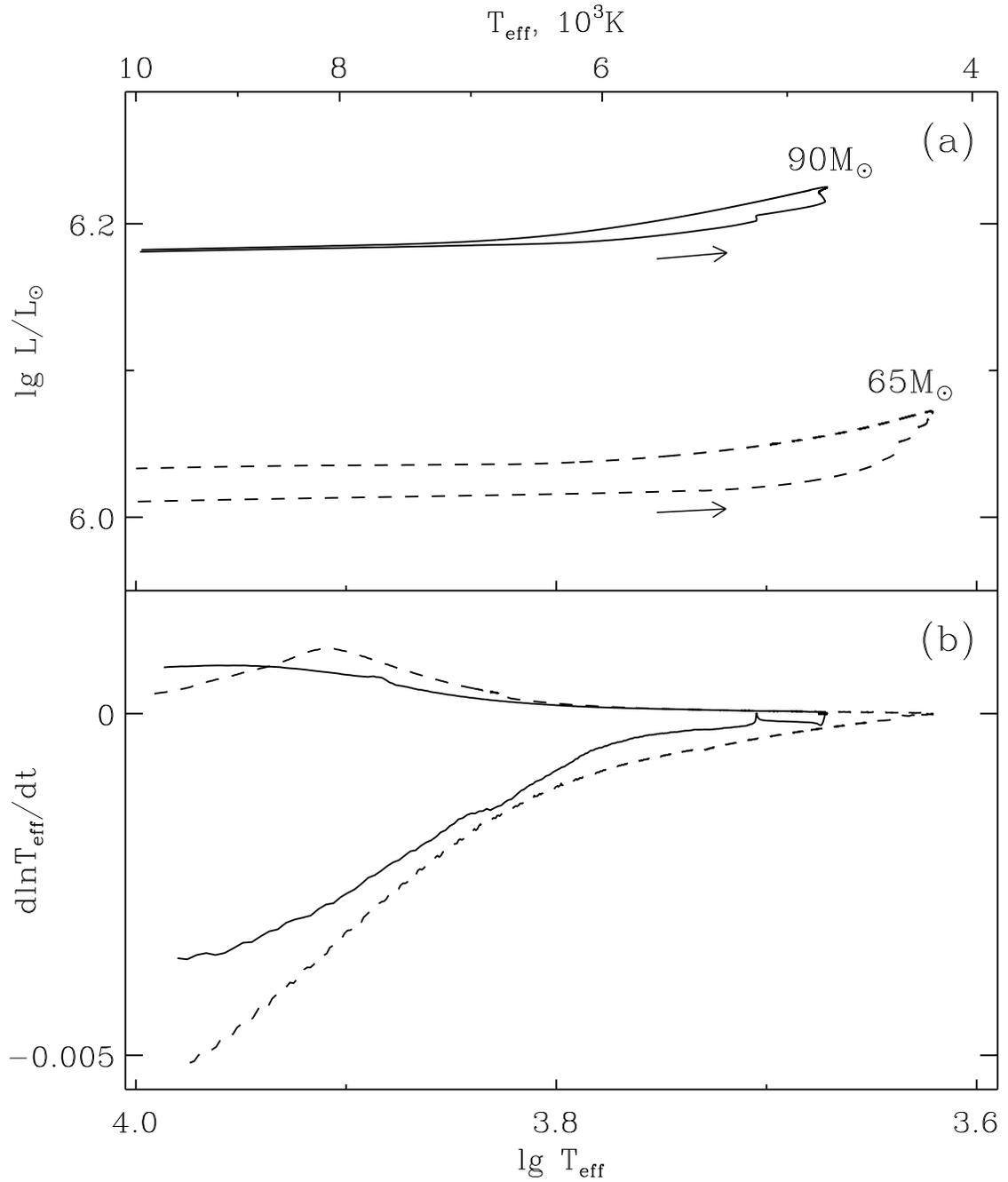}}
\caption{(a) -- Parts of the evolutionary tracks of stars with initial mass
$\mzams=90M_\odot$ (solid line) и $\mzams=65M_\odot$ (dashed line)
during the post--main--sequence core contraction.
Arrows indicate the direction of evolution along the track.
(b) -- The rate of effective temperature change $d\ln\Teff/dt$ along the track.}
\label{fig1}
\end{figure}

Yellow hypergiants are on the short evolutionary stage of gravitational contraction of the
helium core during of which the star evolves along the loop in the Hertzsprung--Russell (HR)
diagram and its effective temperature becomes as low as
$4\times 10^3\K\lesssim\Teff\lesssim 5\times 10^3\K$.
The parts of the evolutionary tracks of stars with effective temperatures $\Teff\le 10^4$K
and initial masses $\mzams=65M_\odot$ and $90M_\odot$ are shown in Fig.~\ref{fig1}(a).
The time of evolution along the shown parts of the tracks ranges from $\approx 10^4$~yr
($\mzams=90M_\odot$) to $\approx 2\times 10^4$~yr ($\mzams=65M_\odot$).

During evolutionary expansion of outer layers when the stellar effective temperature is
$\Teff=10^4$K (the initial point of the tracks in Fig.~\ref{fig1}(a))
the principal energy source is gravitational contraction of
the helium core.
For example, the ratio of central values of the thermonuclear energy generation rate
$\varepsilon_\mathrm{c}$ (these are mostly reactions of the 3--$\alpha$ process)
and the gravitational energy release ranges from
$\alphac = \varepsilon_\mathrm{c}/(-T_\mathrm{c}\partial S_\mathrm{c}/\partial t)\approx 7\times 10^{-3}$
for $\mzams=65M_\odot$ to $\alphac\approx 0.05$ for $\mzams=90M_\odot$.
Here $T_\mathrm{c}$ and $S_\mathrm{c}$ are the temperature and specific entropy
in the center of the star.
This ratio gradually increases with time and at the turning point of the track is
$\alphac\approx 10^2$ independently of $\mzams$.
At ending points of the tracks shown in Fig.~\ref{fig1}(a) this ratio is
$\alphac\approx 300$  for $\mzams=65M_\odot$ and
$\alphac\approx 10^3$  for $\mzams=90M_\odot$.

Gradual decrease of the role of gravitational contraction in the total stellar
energy production leads to deceleration of the evolutionary movement along the track.
This effect is illustrated in Fig~\ref{fig1}(b), where for tracks in Fig~\ref{fig1}(a)
we give the plots of the rate of effective temperature change $d\ln\Teff/dt$.
For example, at $\Teff\approx 7\cdot 10^3$K the rate of the redward evolution
in the HR diagram about five times of that in the opposite direction at the
same effective temperature.
Thus, the probability to observe the yellow hypergiant with increasing effective temperature
is significantly higher than that to observe the hypergiant with expanding
outer layers.
Nevertheless in our study we consider the pulsation properties of hypergiants with
effective temperatures $\Teff\le 10^4$K independently of direction of their evolution
in the HR diagram.

\subsection{Pulsations of yellow hypergiants}

The study of self--exciting stellar oscillations implies solution of the
Cauchy problem for equations (\ref{motcon}) -- (\ref{enereq2}) with initial
conditions corresponding to the hydrostatic and thermal ($\nabla\cdot\mathbf{F}=0$)
equilibrium.
To this end we used some stellar models of preliminary computed evolutionary sequnces.
Stability of the star against radial oscillations is determined from
integration of the equations of hydrodynamics with respect to time $t$.
In the case of pulsational instability the solution describes the exponetially growing
oscillation amplitude with following transition to the limit cycle oscillations.
The main quantities characterizing the limit cycle are the period $\Pi$
and the amplitude of the radial displacement of the outer boundary $\Delta R$
expressed in units of the equilibrium stellar radius $R$.

In Fig.~\ref{fig2} the plots of $\Delta R/R$ and $\Pi$ as a function of $\Teff$
are shown for the evolutionary sequence of models with initial mass $\mzams=65M_\odot$.
As in the HR diagram the effective temperature increases from right to left.
The dashed lines correspond to the evolutionary expansion of outer layers of the star,
that is to decrease of the effective temperature, whereas solid lines correspond
to their contraction.

Radial oscillations of the hypergiant at the initial point of the evolutionary track
($\Teff=10^4$K) are driven by the layers of the iron Z--bump ($T\sim 2\times 10^5$K)
and by helium ionization zones locating closer to the surface.
The role of the iron Z--bump in pulsational instability decreases with
decreasing effective temperature because of diminution of the relative radius $r/R$
and the amplitude of radial displacement $\Delta r$ in these layers.
That is why the evolutionary decrease of the effective temperature is accompanied
by decreasing pulsation amplitude $\Delta R/R$.
This conclusion is illustrated in Fig.~\ref{fig3} where for two hydrodynamical models
with effective temperatures $\Teff=10^4$K and $\Teff=6800$K
we show the radial dependencies of the mechanical work over a closed thermodynamic cycle:
\begin{equation}
W_j = \frac{\Delta M_j}{|W|} \oint P_j dV_j .
\end{equation}
Here $\Delta M_j$ is the mass of the $j$--th Lagrangean interval,
$P_j$ is the sum of the thermodynamic pressure and turbulent pressure,
$V_j$ is the specific volume,
\begin{equation}
|W| = \sum\limits_{j=1}^N \Delta M_j \oint \bigl\vert P_j dV_j\bigr\vert
\end{equation}
is the normalizing coefficient,
$N$ is the number of Lagrangean zones in the hydrodynamical model.
Depending on the stellar structure the hydrodynamic computations were
done with $500\le N\le 1000$.

As seen in Fig.~\ref{fig3} the evolutionary decrease of effective temperature from
$10^4$K to 6800K is accompanied by decrease of the relative radius of the layer with
maximum mechanical work $W_j$ at the iron Z--bump
from $r/R\approx 0.3$ to $r/R\approx 0.15$.
At the same time the contribution of the iron Z--bump zone into the positive mechanical
work decreases from 0.43 to 0.1, so that near the boundary of the pulsational instability
radial pulsations are driven mostly in the helium ionization zones.
Here we have to note that the plots of the mechanical work $W_j$ in Fig.~\ref{fig3}
are shown as a function of the equilibrium relative radius $r/R$, whereas
the mechanical work of a group of mass zones is determined from integration
with respect to mass coordinate $M_r$.

Evolutionary expansion of the hypergiant outer layers is accompanied by
increase of convection in the helium ionization zones, so that for $\Teff\lesssim 6700$K
the radial oscillations decay due to large decrease of the radiative flux
in ionization zones.
Recommencement of radial oscillations becomes possible after the turn of the evolutionary
track  when the effective temperature rises up to $\Teff\gtrsim 7300$K.
Pulsational properties of the hypergiant during evolutionary contraction
of outer layers are illustrated in Fig.~\ref{fig4}, where radial dependencies of the mechanical work
over a closed thermodynamic cycle are shown for hydrodynamical models with $\Teff=7400$K and $10^4$K.

The star of effective temperature $\Teff=7400$K  with contracting outer layers
locates near the boundary of the pulsational instability region.
In comparison with the hypergiant of the same effective temperature but
on the preceding evolutionary stage of expanding outer layers
(the luminosity and the surface mass fraction of helium are
$L=1.033\times 10^6L_\odot$ and $Y_\mathrm{s}=0.74$)
the star has a somewhat higher luminosity and perceptibly higher
surface abundance of helium:
$L=1.085\times 10^6L_\odot$, $Y_\mathrm{s}=0.88$.
The higher helium abundance leads to a larger opacity of stellar material and
to stronger convection
in the helium ionization zones, so that the boundary of the pulsation instability
moves to somewhat higher effective temperatures in comparison with hypergiants
undergoing expansion of outer layers.

Pulsational instability of the hypergiant during evolutionary contraction of its
outer layers is due to the $\kappa$--mechanism  in the helium ionization zones.
The radial dependence of the mechanical work over a closed thermodynamic cycle
for the model near the boundary of instability region is shown in Fig.~\ref{fig4}
by the dashed line.
Two prominent positive maxima on the plot correspond to the zones of the first
ionization and
second ionization of helim, whereas the contribution of the iron Z--bump zone
into the positive mechanical work does not exceed 6\%.
During further evolution with increasing effective temperature the role of
the iron Z--bump zone remains insignificant and at $\Teff=10^4$K its contribution
into the total positive mechanical work remain less than 8\%.

Of greatest interest in comparison with observations is the pulsation period $\Pi$.
During the evolutionary expansion of outer layers in the hypergiant with initial mass
$\mzams=65M_\odot$ the radial pulsations are due to instability of the
fundamental mode and as seen in Fig.~\ref{fig2}(b) the period
gradually increases with decreasing $\Teff$ up to the instability boundary.
During movement along the track in the opposite direction
the radial oscillations resume at effective temperature $\Teff=7400$K.
For $\Teff < 7700$K the star pulsates in the first overtone and
the period of the contracting hypergiant is significantly shorter than that
of the expanding hypergiant with the same effective temperature.
Transition of radial pulsations from the first overtone to the fundamental mode takes place
at effective temperature $\Teff\approx 7700$K and
is revealed by a jump in plots of the surface amplitude $\Delta R/R$ and period $\Pi$
(see Fig.~\ref{fig2}).

To compare the hypergiant pulsation properties at evolutionary stages
of expansion and contraction
we give in Fig.~\ref{fig5} the plots of the gas flow velocity at the outer boundary $U$
and the light in V band $m_\mathrm{V}$ for two hydrodynamical models with initial mass
$\mzams=65M_\odot$ and effective temperature $\Teff=7500$K.
The significant difference in amplitudes of velocity and light curves
is due to the fact that evolutionary contracting hypergiant pulsates in the first obertone,
whereas during evolutionary expansion the hypergiant pulsates in
the fundamental mode.
However at effective temperatures $\Teff > 7700$K
the difference in pulsational properties becomes insignificant (see Fig.~\ref{fig6})
because on both evolutionary stages the hypergiant pulsates in the fundamental mode.

Above we discussed results of hydrodynamic calculations for stars with initial mass
$\mzams=65M_\odot$.
Hypergiants with larger initial mass have a higher helium abundance in their
outer layers because of the more massive convective core and the stronger stellar
wind during the main--sequence evolutionary stage.
Therefore in more massive hypergiants with $\Teff < 10^4$K the convection
in helium ionization zones is stronger and the boundary of the pulsation instability
region corresponds to higher effective temperatures.
For example, in stars with initial mass $\mzams=90M_\odot$
the radial pulsations cease during evolutionary expansion of outer layers
at effective temperature $\Teff\approx 7200$K and resume during evolutionary contraction
at $\Teff\approx 7600$K.

In Fig.~\ref{fig7} the amplitude of the radial displacement of the outer boundary
$\Delta R/R$ and the pulsation period $\Pi$ are shown as a function of effective
temperature for hypergiants with initial mass $\mzams=90M_\odot$.
With approaching the boundary of pulsational instability
during the evolutionary expansion of outer layers
the mode of radial oscillations changes and
hypergiants with effective temperature $\Teff < 7800$K pulsate in the first overtone.
Therefore, the maximum period of radially pulsating hypergiant with initial mass
$\mzams=90M_\odot$ does not exceed 150~day.
During movement along the track in the opposite direction the radial oscillations
resume in the first overtone at effective temperature $\Teff\approx 7700$K
and at $\Teff\gtrsim 8400$K the star oscillates in the fundamental mode.

The main properties of some hydrodynamic models of radially pulsating hypergiants
with initial masses $\mzams = 65M_\odot$ and $90M_\odot$ are summarized in the table.
All models are represented in four sequences depending on the initial stellar mass
$\mzams$ and direction of movement in the HR diagram.
The stellar mass $M$, the luminosity $L$ and the surface fractional mass abundance of
hydrogen $X_\mathrm{s}$ approximately do not change within each sequence and therefore
are given only for the first model of the sequence.
The luminosity $L$ and effective temperature $\Teff$ correspond to the
initial hydrostatic equilibrium,
$\Pi$ and $Q$ are the pulsation period and pulsation constant in days,
$k$ is the order of the pulsation mode ($k=0$ for the fundamental mode
and $k=1$ for the first overtone).
The rate of period change $\dot\Pi/\Pi = d\ln\Pi/dt$ was evaluated
from the second--order numerical differentiation with respect to
age of the stellar evolution model used as initial conditions in
hydrodynamic calculations.
Thus, to evaluate the rate of period change we had to compute two
additional hydrodynamical models.
In the table we give only most reliable estimates of this quantitiy.
The amplitude of the radial displacement of the outer boundary $\Delta R$
expressed in units of the initial equilibrium radius $R$ and the amplitude
of the radial velocity at the outer boundary $\Delta U$
are given in last two columns of the table.

\section{Conclusions}

Results of hydrodynamic calculations described in the present paper
allow us to conclude that radial oscillations is rather a short--term
phenomenon in the life of yellow hypergiants.
For example, hypergiants with initial masses $65M_\odot\le\mzams\le 90M_\odot$
and effective temperatures $\Teff\le 10^4$K are stable against radial oscillations
during $\approx 95$\% of the time needed to move along the loop
of the evolutionary track.
The stability against radial oscillations is due to the strong convection
in helium ionization zones which suppresses the $\kappa$--mechanism of pulsation
instability, whereas the zone of the iron Z--bump locates to deep in the
stellar envelope and does not play any perceptible role.
Therefore, radial pulsations of yellow hypergiants take place only in the
beginning and in the end of this evolutionary stage.

Estimates of the upper limit of period of radially pulsating hypergiants
obtained in the present study can be applied to interpretaion of some
variable stars.
For example, semi--regular light variations of $\rho$~Cas
with period of $300~\mathrm{day}\lesssim\Pi\lesssim 500~\mathrm{day}$
definitely cannot be explained in terms of radial stellar oscillations.
Another example is the pulsating variable star V1427~Aql
with uncertain evolutionary status.
In some studies this variable is discussed as the post--AGB stars, whereas in other
works as a yellow hypergiant (see, for discussion, Le Coroller et al. 2003).
Multicolor photometric observations of V1427~Aql carried out during eight years
reveal the presence of light variations with periods of 130 and 200 day,
the star moving blueward in the HR diagram (Archipova et al. 2009).
Results of our computations allow us to conclude that V1427~Aql is rather
the post--AGB star because assumption on yellow hypergiant leads to
following contradictions.
First, the mean effective temperature of this star is $\Teff = 6750$K
(Kipper 2008) and corresponds to the region of pulsational stability.
Second, the period of light changes $\approx 200$~day significantly exceeds
its upper limit for radially pulsating yellow hypergiants
evolving blueward in the HR diagram.

Discussed above results of hydrodynamic calculations deal with massive stars
with effective temperatures $\Teff\le 10^4$K.
In the present study we carried out hydrodynamic computations for some
models with higher effective temperatures corresponding to the both
earlier and later stages of stellar evolution.
In the first case pulsational instability allows us to conclude that
radial oscillations of main--sequence stars with initial mass $\mzams > 60M_\odot$
(Fadeyev 2011) exist during the post--main--sequence stage and decay only
at effective temperature $\Teff\approx 7000$K due to convection.
In the second case radial oscillations arise at the final stage of the yellow
hypergiant and do not decay up to exhaustion of helium in the stellar core.
With increasing effective temperature the hypergiant becomes the LBV--star
radial pulsations of which are responsible for the microvariability
(Fadeyev 2010) and then transforms into the Wolf--Rayet star which is also
pulsationally unstable (Fadeyev 2007, 2008a, 2008b).

\newpage
\section*{REFERENCES}

\begin{enumerate}

\item A. Arellano Ferro, MNRAS \textbf{216}, 571 (1985).

\item V.P. Arkhipova, V.F. Esipov, N.P. Ikonnikova, G.V. Komissarova, A.M. Tatarnikov, B.F. Yudin,
      Pis'ma Astron. Zh., \textbf{35}, 846 (2009)
      [Astron. Lett., \textbf{35}, 764 (2009)].

\item E. B\"ohm--Vitense, Zeitshrift f\"ur Astrophys. \textbf{46}, 108 (1958).

\item H. Le Coroller, A. L\'ebre, D. Gillet, et al., Astron. Astrophys. \textbf{400}, 613 (2003).

\item Yu.A. Fadeyev, Pis'ma Astron. Zh., \textbf{33}, 775 (2007)
      [Astron. Lett. \textbf{33}, 692 (2007)].

\item Yu.A. Fadeyev, Astron. Zh. \textbf{85}, 716 (2008a)
      [Astron. Rep. \textbf{54}, 645 (2008)].

\item Yu.A. Fadeyev, Pis'ma Astron. Zh., \textbf{34}, 854 (2008b)
      [Astron. Lett. \textbf{34}, 772 (2008)].

\item Yu.A. Fadeyev, Pis'ma Astron. Zh., \textbf{36}, 380 (2010)
      [Astron. Lett. \textbf{36}, 362 (2010)].

\item Yu.A. Fadeyev, Pis'ma Astron. Zh., \textbf{37}, 13 (2011)
      [Astron. Lett. \textbf{37}, 11 (2011)].

\item M. W. Feast and A. D. Thackeray, MNRAS \textbf{116}, 41 (1956).

\item N. Gorlova, A. Lobel, A.J. Burgasser, et al.), Astrophys.J \textbf{651}, 1130 (2006).

\item C. de Jager, \textit{The Brightest Stars} (Reidel, Dordrecht, 1980; Mir, Moscow, 1984).

\item C. de Jager, Astron. Astrophys. Rev. \textbf{8}, 145 (1998).

\item C. de Jager and H. Nieuwenhuijzen, MNRAS \textbf{290}, L50 (1997).

\item C. de Jager, A. Lobel, H. Nieuwenhuijzen, et al., MNRAS \textbf{327}, 452 (2001).

\item T. Kipper, Baltic Astron. \textbf{17}, 87 (2008).

\item R. Kuhfuss, Astron. Astrophys. \textbf{160}, 116 (1986).

\item N. Langer, C.A. Norman, A. de Koter, et al., Astron. Astrophys. \textbf{475}, L19 (2007).

\item A. Lobel, C. de Jager, H. Nieuwenhuijzen, et al., Astron. Astrophys \textbf{291}, 226 (1994).

\item A. Lobel, G. Israelian, C. de Jager, et al., Astron. Astrophys. \textbf{330}, 659 (1998).

\item G. Meynet, A. Maeder, G. Schaller, et al., Astron. Astrophys. Suppl. \textbf{103}, 97 (1994).

\item E.A. Olivier and P.R. Wood, MNRAS \textbf{362}, 1396 (2005).

\item R.D. Richtmyer and K.W. Morton, \textit{Difference Methods for Initial--Value Problems}
      (2nd ed., Interscience, 1967; Mir, Moscow, 1972).

\item Y. Sheffer and D. L. Lambert, PASP \textbf{98}, 914 (1986).

\item R. Smolec and P. Moskalik, Acta Astron. \textbf{58}, 193 (2008).

\item R.B. Stothers,  Astrophys. J. \textbf{589}, 960 (2003).
  
\item R.B. Stothers and C.--W. Chin, Astrophys. J. \textbf{468}, 842 (1996).

\item R.B. Stothers and C.--W. Chin, Astrophys. J. \textbf{560}, 934 (2001).

\item E. Zsoldos and J. B. Percy,  Astron. Astrophys. \textbf{246}, 441 (1991).

\item G. Wuchterl and M.U. Feuchtinger, Astron. Astrophys. \textbf{340}, 419 (1998).

\end{enumerate}

\newpage
\centerline{Hydrodynamic models of yellow hypergiants}
\begin{table}[h]
\begin{tabular}{r|l|r|r|r|r|r|r|r|r|r}
\hline
 $\mzams/M_\odot$ & $M/M_\odot$ & $L/L_\odot$, & $X_\mathrm{s}$ & $\Teff$, & $\Pi$, & $Q$, & $k$ & $\dot\Pi/\Pi$ & $\Delta R/R$ & $\Delta U$,  \\
                  &             & $10^6$       &                & $10^3$K  & day    & day  &     & $10^{-5}~\mathrm{day}^{-1}$     &              & km/s         \\
\hline
 65 & 32.9 & 1.03 & 0.24 & 10   &  63 & 0.0587 & 0 &  3.6  & 0.65 & 140  \\
    &      &      &      &  9   &  82 & 0.0551 & 0 &  3.6  & 0.28 &  79  \\
    &      &      &      &  8   & 115 & 0.0541 & 0 &  2.4  & 0.33 &  84  \\
    &      &      &      &  7   & 168 & 0.0525 & 0 &  1.3  & 0.33 &  77  \\
    &      &      &      &  6.8 & 180 & 0.0516 & 0 &       & 0.28 &  60  \\[4pt]
    & 28.6 & 1.08 & 0.10 &  7.5 & 110 & 0.0380 & 1 & -0.44 & 0.08 &  25  \\
    &      &      &      &  8   & 124 & 0.0523 & 0 & -0.60 & 0.37 &  92  \\
    &      &      &      &  9   &  87 & 0.0524 & 0 & -0.60 & 0.31 &  86  \\
    &      &      &      & 10   &  64 & 0.0525 & 0 &       & 0.28 &  87  \\[4pt]
 90 & 42.2 & 1.52 & 0.16 & 10   &  74 & 0.0584 & 0 &       & 0.42 & 110  \\
    &      &      &      &  9   & 100 & 0.0559 & 0 &  2.5  & 0.32 &  91  \\
    &      &      &      &  8   & 143 & 0.0561 & 0 &       & 0.39 & 101  \\
    &      &      &      &  7.5 & 124 & 0.0402 & 1 &  1.5  & 0.09 &  27  \\[4pt]
    & 38.4 & 1.52 & 0.08 &  7.7 & 123 & 0.0413 & 1 &       & 0.07 &  19  \\
    &      &      &      &  8   & 111 & 0.0421 & 1 & -0.44 & 0.14 &  49  \\
    &      &      &      &  9   & 103 & 0.0553 & 0 & -0.63 & 0.36 & 101  \\
    &      &      &      & 10   &  76 & 0.0563 & 0 &       & 0.31 & 100  \\
\hline
\end{tabular}
\end{table}
\clearpage

\newpage
\section*{FIGURE CAPTIONS}

\begin{figure}
\centerline{\includegraphics[width=15cm]{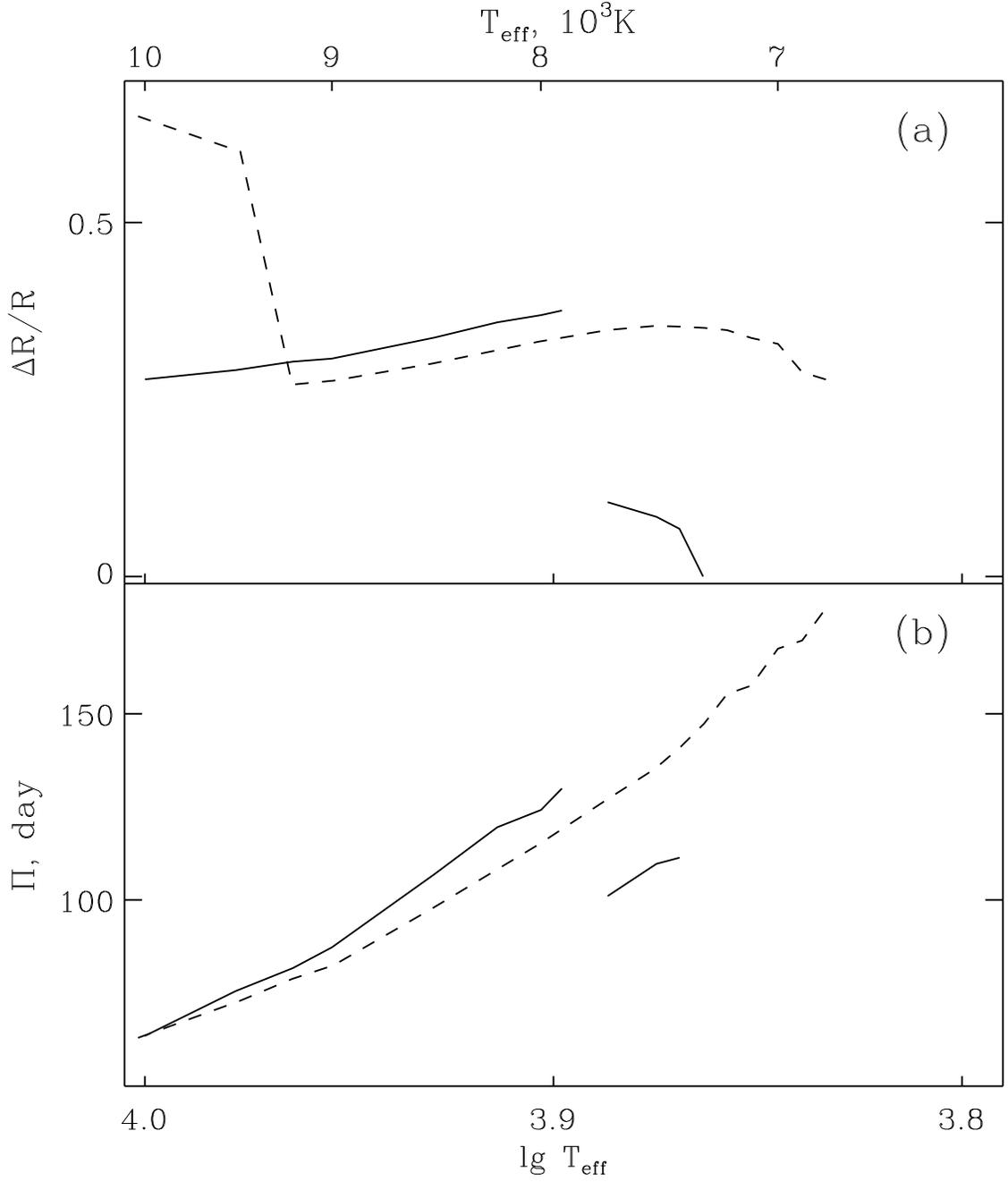}}
\caption{The amplitude of the radial displacement of the outer boundary $\Delta R/R$
(a) and the period of radial pulsations $\Pi$ (b)
as a function of the effective temperature $\Teff$
for models of the evolutionary sequence $\mzams=65M_\odot$ 
Dashed lines correspond to the evolutionary expansion of outer layers of the star
and solid lines correspond to their contraction.}
\label{fig2}
\end{figure}

\begin{figure}
\centerline{\includegraphics[width=15cm]{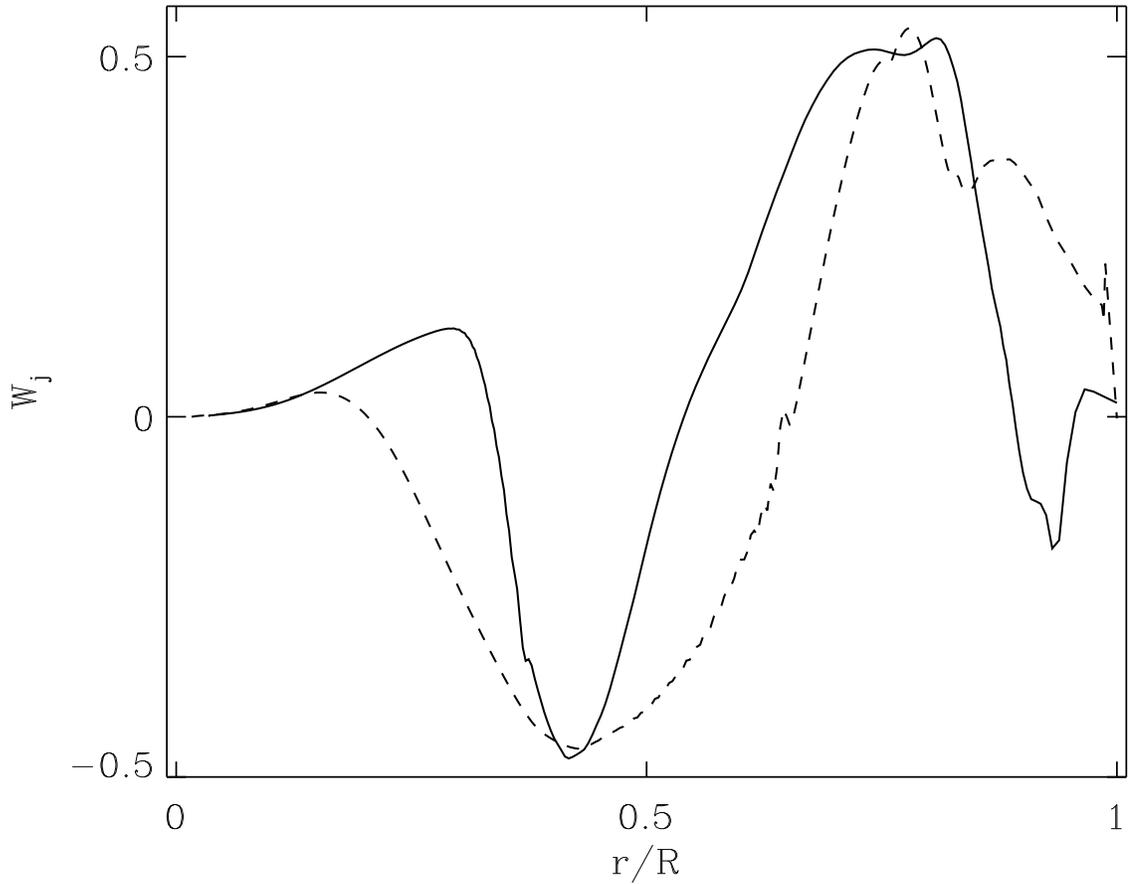}}
\caption{The normalized mechanical work over a closed thermodynamic cycle $W_j$
as a function of the relative equilibrium radius $r/R$
at the evolutionary stage of expansion of outer layers of the star
$\mzams=65M_\odot$
at effective temperature $\Teff=10^4$K (solid line) and
$\Teff=6800$K (dashed line).}
\label{fig3}
\end{figure}

\begin{figure}
\centerline{\includegraphics[width=15cm]{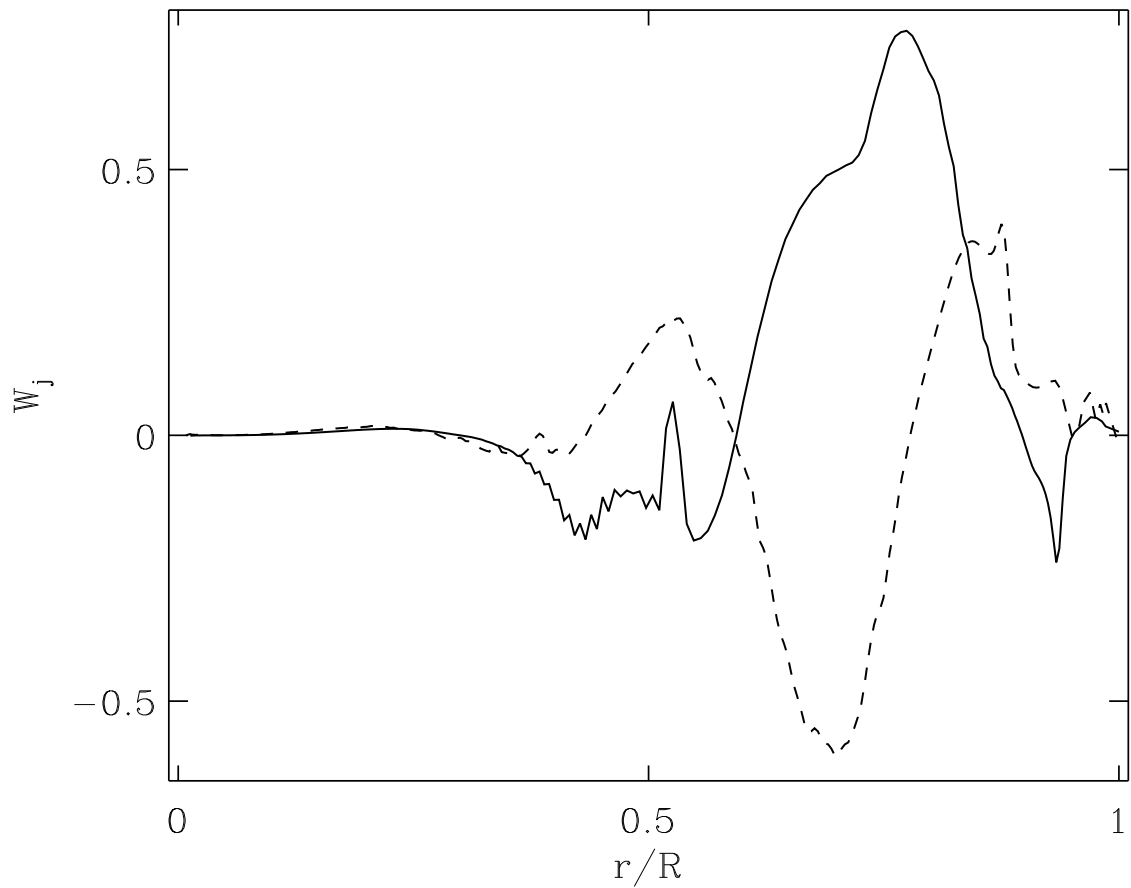}}
\caption{Same as in Fig.~\ref{fig3} but for the evolutionary stage of
contraction of outer layers of the star with $\Teff=7400$K (dashed line) and
$\Teff=10^4$K (solid line).}
\label{fig4}
\end{figure}

\begin{figure}
\centerline{\includegraphics[width=15cm]{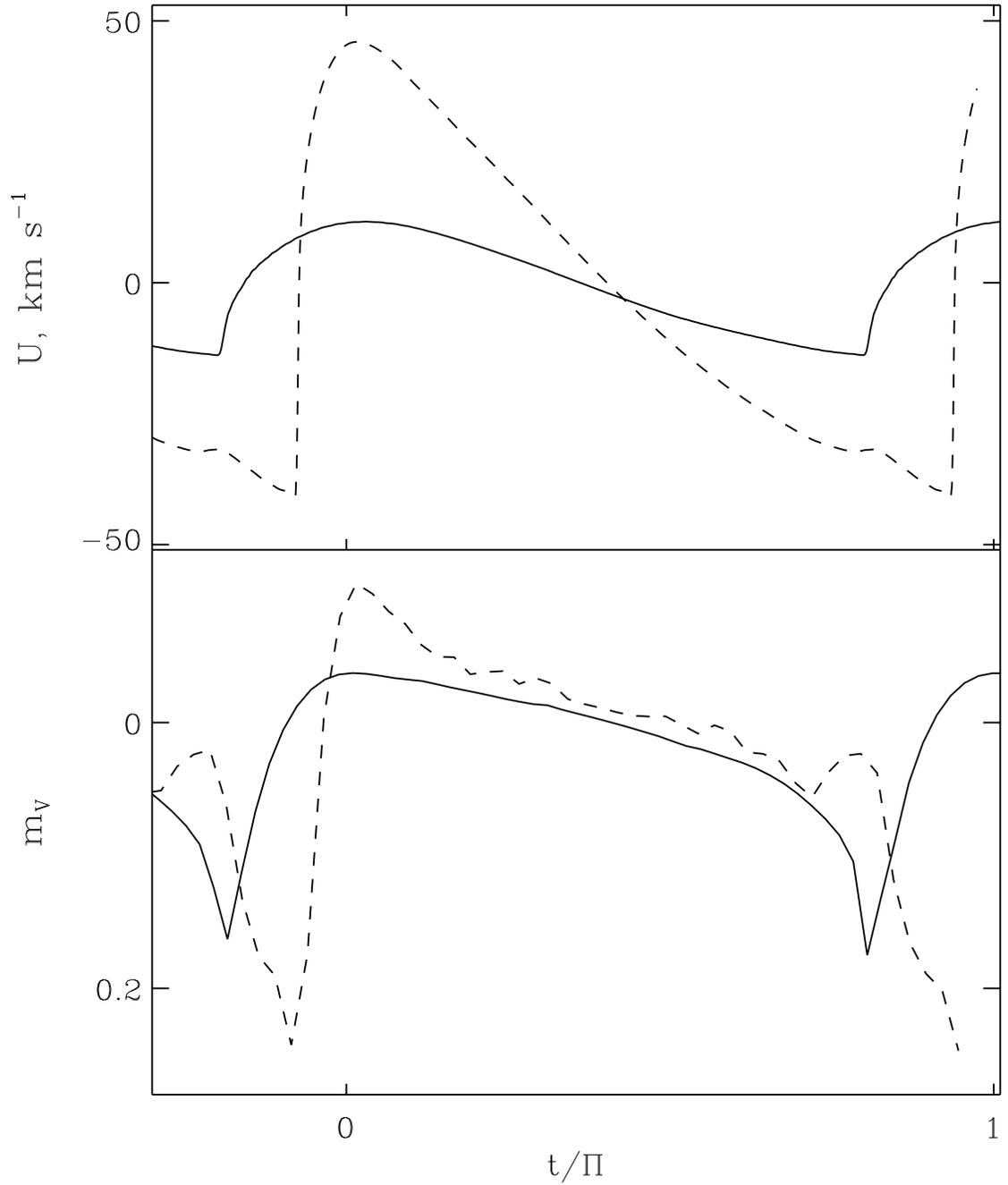}}
\caption{The gas flow velocity at the outer boundary $U$ (a) and the light in V band
$m_\mathrm{V}$ (b)
of hydrodynamic models of the hypergiant with initial mass $\mzams=65M_\odot$ 
and effective temperature $\Teff=7500$K.
Solid and dashed lines correspond to evolutionary stages of contraction and
expansion, respectively.}
\label{fig5}
\end{figure}

\begin{figure}
\centerline{\includegraphics[width=15cm]{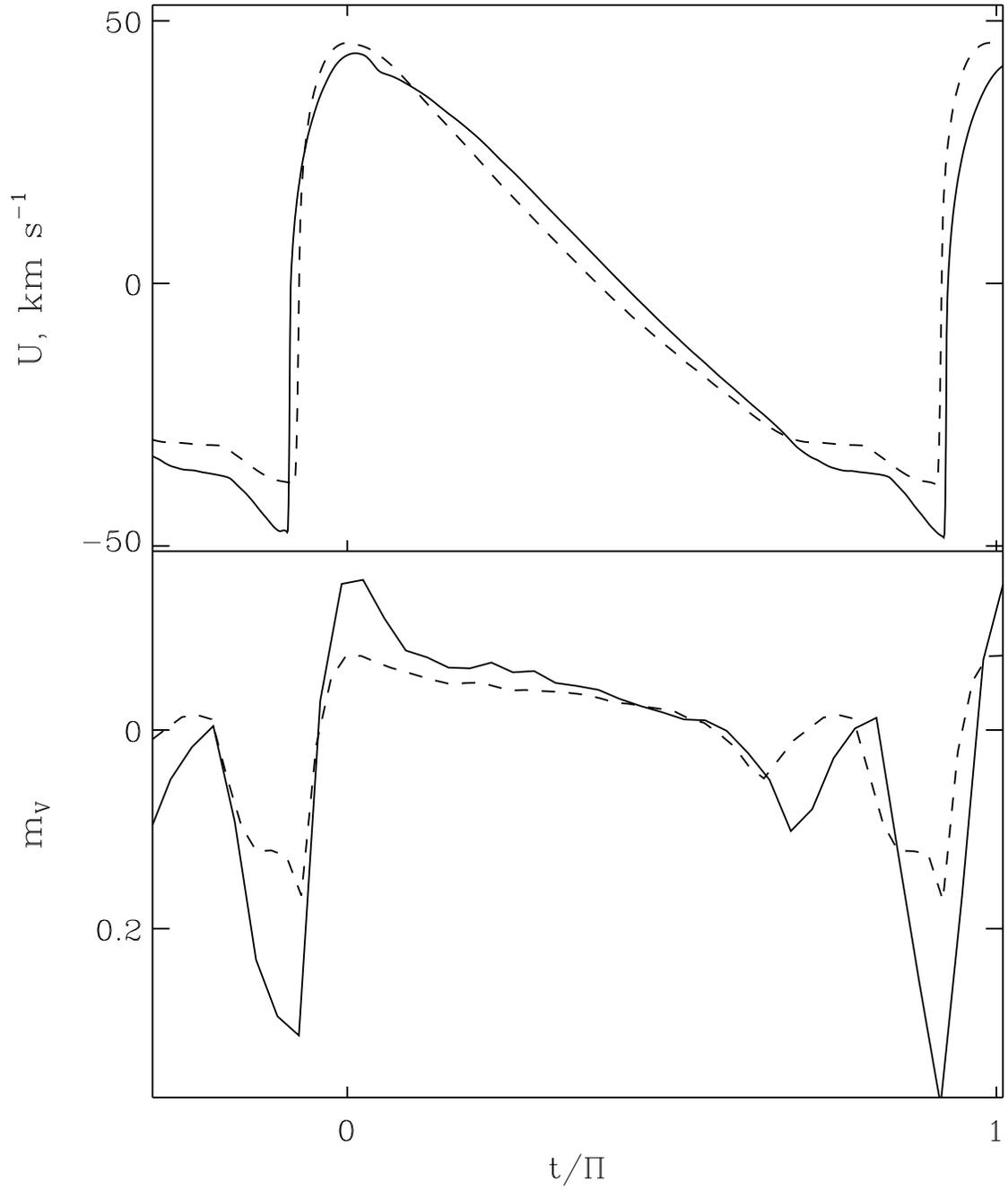}}
\caption{Same as in Fig.~\ref{fig5} but for hydrodynamical models
of the hypergiant with effective temperature $\Teff=8000$K.}
\label{fig6}
\end{figure}

\begin{figure}
\centerline{\includegraphics[width=15cm]{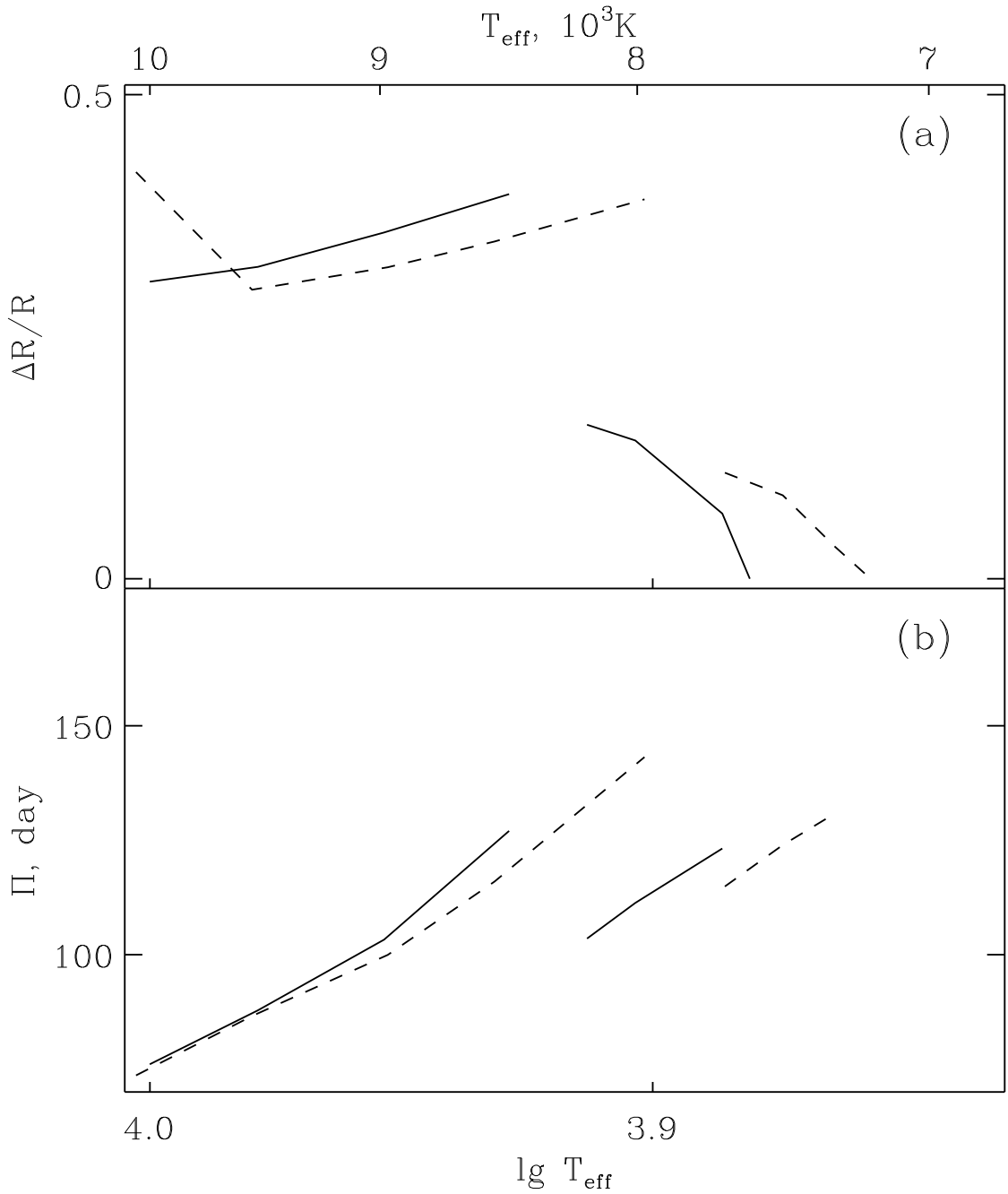}}
\caption{Same as in Fig.~\ref{fig2} but for hypergiants with
initial mass $\mzams=90M_\odot$.}
\label{fig7}
\end{figure}

\begin{itemize}
\item[Fig. 1.]
(a) -- Parts of the evolutionary tracks of stars with initial mass
$\mzams=90M_\odot$ (solid line) and $\mzams=65M_\odot$ (dashed line)
during the post--main--sequence core contraction.
Arrows indicate the direction of evolution along the track.
(b) -- The rate of effective temperature change $d\ln\Teff/dt$ along the track.

\item[Fig. 2.]
The amplitude of the radial displacement of the outer boundary $\Delta R/R$
(a) and the period of radial pulsations $\Pi$ (b)
as a function of the effective temperature $\Teff$
for models of the evolutionary sequence $\mzams=65M_\odot$ 
Dashed lines correspond to the evolutionary expansion of outer layers of the star
and solid lines correspond to their contraction.

\item[Fig. 3.]
The normalized mechanical work over a closed thermodynamic cycle $W_j$
as a function of the relative equilibrium radius $r/R$
at the evolutionary stage of expansion of outer layers of the star
$\mzams=65M_\odot$
at effective temperature $\Teff=10^4$K (solid line) and
$\Teff=6800$K (dashed line).

\item[Fig. 4.]
Same as in Fig.~\ref{fig3} but for the evolutionary stage of
contraction of outer layers of the star with $\Teff=7400$K (dashed line) and
$\Teff=10^4$K (solid line).

\item[Fig. 5.]
The gas flow velocity at the outer boundary $U$ (a) and the light in V band
$m_\mathrm{V}$ (b)
of hydrodynamic models of the hypergiant with initial mass $\mzams=65M_\odot$ 
and effective temperature $\Teff=7500$K.
Solid and dashed lines correspond to evolutionary stages of contraction and
expansion, respectively.

\item[Fig. 6.]
Same as in Fig.~\ref{fig5} but for hydrodynamical models
of the hypergiant with effective temperature $\Teff=8000$K.

\item[Fig. 7.]
Same as in Fig.~\ref{fig2} but for hypergiants with
initial mass $\mzams=90M_\odot$.

\end{itemize}

\newpage

\newpage

\newpage

\newpage

\newpage

\newpage

\newpage

\end{document}